\documentstyle[aps,multicol,prl,epsfig]{revtex}

\begin{document}

\title{On the Modulational Instability of the Nonlinear Schr{\"o}dinger 
Equation with Dissipation}

\author{Z. Rapti$^1$, P.G. Kevrekidis$^1$, D.J. Frantzeskakis$^2$
and B.A. Malomed$^3$}
\address{$^1$ Department of Mathematics and Statistics, University of 
Massachusetts, 
Amherst MA 01003-4515, USA\\
$^{2}$ Department of Physics, University of Athens,Panepistimiopolis, 
Zografos, Athens 15784, Greece \\
$^3$ Department of Interdisciplinary Studies, Faculty of Engineering, 
Tel Aviv University, Tel Aviv 69978, Israel }
%\\
%$^4$     Department of Applied Mathematics, University of Colorado at 
%Boulder, Boulder, 
%Colorado 80309, USA  }

\maketitle

\begin{abstract}
The modulational instability (MI) of spatially uniform states in the nonlinear
Schr{\"{o}}dinger (NLS) equation is examined in the presence of higher-order
dissipation. The study is motivated by results on the effects of
three-body recombination in Bose-Einstein condensates (BECs), as well
as by the important recent work of Segur {\it et al.} on the effects
of linear damping in NLS settings. 
%It was 
We show how
the presence of even the weakest possible dissipation suppresses the
instability on a longer time scale. However, on a shorter scale, the
instability growth may take place, and a corresponding generalization of the MI
criterion is developed. The analytical results are corroborated by numerical
simulations. The method is valid for any power-law dissipation form,
including the constant dissipation as a special case.
\end{abstract}

\vspace{2mm}

\begin{multicols}{2}

\section{Introduction}

The nonlinear Schr{\"{o}}dinger (NLS) equation is well known to be
a generic model to describe evolution of envelope waves in
nonlinear dispersive media. This equation applies to
electromagnetic fields in optical fibers \cite {hasegawa}, deep
water waves \cite{benjamin}, Langmuir waves \cite{Berge'} and
other types of perturbations in plasmas, such as electron-acoustic
\cite {electron-acoustic} waves and waves in dusty plasmas
\cite{dust}, macroscopic wave functions of Bose-Einstein
condensates (BECs) in dilute alkali vapors \cite{book} [in the latter
case, it is usually called the Gross-Pitaevskii (GP) equation],
and many other physical systems. The NLS typically 
appears in the Hamiltonian form
\begin{equation}
iu_{t}=-u_{xx}+b|u|^{2}u,  \label{deq0}
\end{equation}
where $u$ is a complex envelope field and $b$ represents the
strength of the nonlinearity (for instance, in the BEC context,
$u$ is the mean-field wave function and $b$ is proportional to the
s-wave scattering length \cite{book}). The cases $b>0$ and $b<0$
correspond, respectively, to the self-focusing and self-defocusing
NLS equation.

In many physically relevant cases, dissipation, which is
ignored in Eq. (\ref{deq0}), is not negligible. In particular, the
fiber loss gives rise to a linear dissipative term in the NLS
equation for optical fibers \cite {hasegawa,lisak,Karlsson}. Dissipative
phenomena have also been studied recently in the context of
two-dimensional water waves \cite{segur,segur1} in order to
explain experimental observations in water tanks. In this setting also,
the dissipation is accounted for by a linear term. In fact,
it was the viewpoint of \cite{segur,segur1} on the effects of 
damping that chiefly motivated the present work. 
Furthermore, it has been argued that dissipative phenomena are
quite relevant in the context of BECs \cite{3body,nature}.
However, in the latter case, the loss arises from higher-order
processes (such as three-body recombination, corresponding to a
quintic dissipation term in the GP equation \cite{3body}) or from
dynamically-induced losses (such as crossing the Feshbach
resonance \cite {nature,inouye}). Thus, it is necessary to study
models of the NLS/GP type with more complex forms of dissipative
terms. Such terms were also introduced in recent studies of 
collapse in BECs with negative scattering length \cite {kagan} and
evolution of matter-wave soliton trains \cite{leung}.

Higher-order loss -- in the form of a term accounting for {\it five-photon}
absorption -- was also found to play a crucially important role (along with
a cubic term that takes into regard the stimulated Raman scattering) as a
mechanism arresting spatiotemporal collapse in self-focusing of light in a
planar silica waveguide with the Kerr nonlinearity \cite{Shimshon}. In this
case, the high order of the absorption is explained by the fact that the
ratio of the resonant absorption frequency to the carrier frequency of the
photons is close to $5$.

One of the most important dynamical features of the NLS equation
is the modulational instability (MI) of spatially uniform, alias
continuous-wave (CW), solutions. The onset of MI leads to the 
formation of solitary waves and coherent structures in all of the
above settings \cite
{hasegawa,benjamin,Berge',electron-acoustic,dust,Karlsson,segur,sulem,modin}.
MI is a property of special interest in the above-mentioned
dissipative models. 
In particular, as it has been recently argued 
\cite{segur,segur1}  in the context of water waves, dissipation
may lead to the 
complete asymptotic 
suppression of the instability and of the formation of nonlinear 
waves.

%In particular, in the case of the linear loss,
%the MI-induced growth law for the perturbation amplitude in the
%NLS equation linearized around the CW solution was studied in
%detail in \cite{hasegawa,lisak} and more recently in \cite{Karlsson}, 
%with application to optical fibers. In the context of the two-dimensional NLS
%equation with the linear dissipative term modeling water waves, it
%has been recently shown in Refs. \cite{segur,segur1} that
%dissipation leads to complete asymptotic (at $t\rightarrow \infty
%$) suppression of the MI.

The aim of the present work is to examine the MI in the context of the NLS
equation with {\em nonlinear dissipation}. Motivated by the model for BECs
developed in Ref. \cite{3body}, but trying to keep the formulation as
general as possible, we thus consider the damped NLS equation of the form:
\begin{equation}
iu_{t}=-u_{xx}+b|u|^{2}u+(c_{1}-ic_{2})|u|^{2k}u,  \label{deq1}
\end{equation}
where $b$ and $c_{1}$ are real parameters. The constant $c_{2}$ is real and
positive (in order for the corresponding term to describe loss, rather than
gain). The value $k=0$ brings one back to the much studied case of the
linear dissipation \cite{hasegawa,lisak,Karlsson,segur}, while $k=2$
corresponds to the quintic loss, accounting for the above-mentioned
three-body recombination collision in BECs \cite{3body}, and $k=5$ -- to the
above-mentioned case of the five-photon absorption, which is believed to be
the mechanism arresting the spatiotemporal collapse in glass \cite{Shimshon}.

Our presentation 
is structured as follows. In Section II, we consider the MI in the
general equation (\ref{deq1}). In Section III, we focus on three special
cases: (i) the (previously studied) linear damping case, 
(ii) $c_{2}=0$ [the Hamiltonian limit of Eq. (\ref{deq1})], and (iii)
the case of purely dissipative nonlinearity, with $c_{1}=b=0$. 
%Comparison
%with the studies of formally similar Ginzburg-Landau equations \cite{Jena}
%suggests that such cases may give rise to dynamical features of special
%interest. 
In Section IV, we complement our analytical findings by numerical
simulations. Section V concludes the paper.

\section{Modulational Instability for the Dissipative NLS Model}

An exact CW (spatially uniform)\ solution to Eq. (\ref{deq1}) is
\begin{eqnarray}
u_{0}(t) &=&\left( F(t)\right)^{-1/(2k)} \exp \left[ \frac{ib}{2c_{2}(1-k)}\left( F(t)\right)^{1-\frac{1}{k}}\right] 
\nonumber
\\
&\times& \exp \left[-\frac{i}{2c_{2}}\left(\frac{c_{1}}{k} \ln F(t)-\beta _{0}\right) \right],  
\label{deq3}
\end{eqnarray}
where $\beta _{0}$ is an arbitrary phase shift,
\begin{equation}
F(t)=B_{0}^{-2k}+2kc_{2}t,  
\label{F}
\end{equation}
and $B_{0}$ is a positive constant. We now look for perturbed solutions of
the NLS equation in the form
\begin{equation}
u(x,t)=u_{0}(t)\left[ 1+\epsilon w(t)\cos (qx)\right] ,  \label{deq2}
\end{equation}
where $w(t)$ and $q$ are the amplitude and constant wavenumber of the
perturbation, while $\epsilon \rightarrow 0$ is an infinitesimal perturbation
parameter. Substituting (\ref{deq2}) into (\ref{deq1}), keeping terms linear
in $\epsilon $, and splitting the perturbation amplitude in its real and
imaginary parts, $w\equiv w_{r}+iw_{i}$, we obtain the following ordinary
differential equations (ODEs)
\begin{eqnarray}
\dot{w}_{r} &=&q^{2}w_{i}+2kc_{2}\left( F(t)\right) ^{-1}w_{r},  \nonumber \\
\dot{w}_{i} &=&-\left( q^{2}+ 2b\left( F(t)^{-1/k}\right) +2kc_{1}\left(
F(t)\right) ^{-1}\right) w_{r},  \label{deq3a}
\end{eqnarray}
where the overdot stands for $d/dt$. These equations can be combined into
the following linear second-order equation for $w_{r}$, %\begin{equation}
%w_r''-\frac{2 k c_2}{B_0^{-2 k}-2 k c_2 t} w_r'+(q^2 (q^2+\frac{2 b}
%{(B_0^{-2 k}-2 k c_2 t)^{\frac{1}{k}}}+\frac{2 c_1 k}{(B_0^{-2 k}-2 k c_2 t)})-
%\frac{4 c_2^2 k^2}{(B_0^{-2 k}-2 k c_2 t)^2}) w_r=0.
%\label{deq4}
%\end{equation}
\begin{equation}
\ddot{w}_{r}+2f(t)\dot{w}_{r}+g(t)w_{r}=0,  \label{deq44}
\end{equation}
where
\begin{eqnarray}
f(t) &\equiv &-kc_{2}/F(t),  \label{f} \\
g(t) &\equiv &-4k^{2}c_{2}^{2}/F(t) \nonumber
\\
     &+& q^{2}\left[ q^{2}+2b/\left( F(t)\right)
^{1/k}+2kc_{1}/F(t)\right] .  \label{g}
\end{eqnarray}
Equation (\ref{deq44}) can be further cast in a more standard
form, $\ddot{y} +h(t)y=0,$ where $h(t)\equiv g(t)+f^{2}(t)$, and
$y\equiv w_{r}(t)\exp \left(
\int_{0}^{t}f(s)ds\right) $. %If we set
%$f(t)=\frac{-k c_2}{B_0^{-2 k}-2 k c_2 t}$ and
%$g(t)=q^2 (q^2+\frac{2 b}{(B_0^{-2 k}-2 k c_2 t)^{1/k}}+\frac{2 c_1 k}{(B_0^{-2 k}-2 k c_2
%t)})-\frac{4 k^2 c_2^2}
%{(B_0^{-2 k}-2 k c_2 t)^2}$,
%then (\ref{deq44}) can be written as
%\begin{equation}
%w_r''+2 f(t) w_r'+g(t) w_r=0.
%\label{deq44}
%\end{equation}
%We thus introduce the transformation $y=w_r(t) e^{\int_0^t f(s) ds}$ to put
% Eq. (\ref{deq44}) into the form,
%\begin{equation}
%y''+h(t) y=0,
%\label{deq45}
%\end{equation}
%where $h(t)=g(t)+f^2(t)$.
%Note that, in the case of the linear dissipation ($k=0$), Eq. (\ref{deq44})
%admits an exact solution in terms of Bessel functions with an imaginary
%index \cite{Karlsson}.

It follows from Eq. (\ref{g}) that, in the case of the self-focusing
nonlinearity ($b$, $c_{1}<0$), the initial condition $\dot{w}_{r}(0)=0$
(which we will adopt below in numerical simulations) leads to $g(0)<0$,
provided that the wavenumber is chosen from the interval
\begin{eqnarray}
q^{2}<q_{{\rm cr}}^{2}&\equiv& -(bB_{0}^{2}+kc_{1}B_{0}^{2k}) \nonumber
\\
&+&\sqrt{(bB_{0}^{2}+kc_{1}B_{0}^{2k})^{2}+4k^{2}c_{2}^{2}B_{0}^{4k}}.
\label{deq47}
\end{eqnarray}
%it follows that
%$h(0)=q^2 (q^2+2 b B_0^2+2 k c_1 B_0^{2 k})-3 k^2 c_2^2 B_0^{4k}<0$.
The condition $g(0)<0$ implies an initial growth of the
perturbation, similar to the case of MI in the usual NLS equation
(\ref{deq0}) \cite {hasegawa,benjamin,modin}. It is interesting to
note that, somewhat counter-intuitively, the interval
(\ref{deq47}) of unstable wavenumbers {\em expands} ($q_{{\rm
cr}}^{2}$ increases) with the increase of the loss constant
$c_{2}$. This result will be verified by means of numerical
simulations in Sec. IV.

However, the perturbation growth can only be transient in the
presence of the loss, which resembles a known situation for the MI
in a conservative medium where the dispersion coefficient changes
its sign in the course of the evolution (this actually pertains to
evolution of an optical signal along the propagation distance in
an inhomogeneous dispersive medium) \cite {variable-dispersion}.
In particular, even when $g(0)$ is negative (i.e., when the condition
(\ref{deq47}) is satisfied and we are in the --transiently-- unstable
regime), as $t$ increases, $g(t)$ increases
monotonically as well and eventually becomes positive. Thus, we
may expect that, until a critical time $t_{0}>0$, the function
$w_{r}$ monotonically grows but then switches from growth
to oscillations. 
%and, in fact, the oscillations' amplitude
%decreases in time. 
This prediction also follows from the consideration
of the asymptotic form of Eqs. (\ref{f}) and (\ref{g}) for
$t\rightarrow \infty $: as the dissipation suppresses $f(t)$ and
the time-dependent part of $g(t)$, in this limit Eq. (\ref{deq44})
gives rise to oscillations of $w_{r}$ at the frequency $q^{2}$
(and, hence, to stable dynamics).

%From this expression one can also evaluate the so-called
%``integrated exponential gain''
%$G=\int_0^{t_0} \sqrt{h(t)} dt$
%$G$, defined in
%\cite{hasegawa}.

%{\it Would it be interesting to analyze the cases $b>0$ and/or $c_1>0$ and
%see what happens? Maybe h is not always increasing.}

\section{Special cases}

In this section we consider, in more detail, particular cases of
special interest: (i) k=0
(ii) $c_{2}=0$, and (iii) $b=c_{1}=0$. 
The first case is the well-known linear damping limit
(given for completeness).
The second
case corresponds to a higher-order version of the conservative NLS
equation [see Eq. (\ref {deq0})], while the last one is a case
of a purely dissipative nonlinearity.
%, that is why both deserve
%special consideration.
%As mentioned above, the cases (ii) and (iii) are quite
%relevant to the BECs.

\subsection{Linear Dissipation}

In the limiting case $k\rightarrow 0$, Eqs. (\ref{deq1})-(\ref{deq44}) lead
to the following results: first, the relevant exponentially
decaying
solution to (\ref{deq1}) reads
\begin{equation}
u_{0}(t)=\exp (-c_{2}t)\exp \left[ i\left( \frac{b}{2c_{2}}\exp
(-2 c_{2}t)-c_{1}t+\beta _{0}\right) \right] .  \label{deq5}
\end{equation}
On the other hand, $w_{r}$ now satisfies the much simpler equation
\begin{equation}
w_{r}^{\prime \prime }+q^{2}\left[ q^{2}+2b\exp (-2 c_{2}t)\right] w_{r}=0,
\label{deq6}
\end{equation}
which can be identified, e.g., with the equation (15.2.1) of \cite{hasegawa}
(see also \cite{lisak,Karlsson}). An explicit, though complicated, solution 
to this equation can be found
using Bessel functions \cite{Karlsson} and the critical time can
be calculated explicitly (see e.g., \cite{hasegawa}).
%:
%\begin{eqnarray}
%w_{r}(t) &=&A_{1}J_{-\frac{iq^{2}}{c_{2}}}\left( \frac{\sqrt{2b\exp
%(2c_{2}t)q^{2}}}{c_{2}}\right) \Gamma \left( 1-\frac{iq^{2}}{c_{2}}\right)
%\nonumber \\
%&+&A_{2}J_{\frac{iq^{2}}{c_{2}}}\left( \frac{\sqrt{2b\exp (2c_{2}t)q^{2}}}{
%c_{2}}\right) \Gamma \left( 1+\frac{iq^{2}}{c_{2}}\right) ,  \label{deq61}
%\end{eqnarray}
%where $J$, is the Bessel function of the first kind and $\Gamma $ is the
%Euler gamma function ($A_{1,2}$ are integration constants). In this case the
%critical time $t_{0}$ can be found explicitly:
%\begin{equation}
%t_{0}=\frac{1}{2|c_{2}|}\log \left( \frac{2|b|}{q^{2}}\right) .  \label{t0}
%\end{equation}

\subsection{The conservative model.}

In the case $c_{2}=0$, an exact CW solution to Eq. (\ref{deq1}) is
\begin{equation}
u_{0}(t)=A_{0}\exp \left[ -i(bA_{0}^{2}+c_{1}A_{0}^{2k})t\right] ,
\label{deq7}
\end{equation}
where $A_{0}$ is an arbitrary constant, and the linearized equation
(\ref{deq44}) assumes the form
\begin{equation}
w_{r}^{\prime \prime }+q^{2}\left[
q^{2}+2(bA_{0}^{2}+kc_{1}A_{0}^{2k}) \right] w_{r}=0,
\label{deq8}
\end{equation}
thus giving rise to the MI criterion,
$q^{2}<-2(bA_{0}^{2}+kc_{1}A_{0}^{2k})$. As it follows from here,
MI is possible if either nonlinearity is self-focusing, i.e.,
either $b<0$ or $c_{1}<0$. If $b<0$ and the higher-order nonlinear terms are 
also focusing, 
then the interval of modulationally unstable  wavenumbers grows. On the 
other hand, if the higher order terms are defocusing , 
then the corresponding interval shrinks and does so faster for higher 
powers (or BEC densities). In particular, the higher-order
self-defocusing nonlinearity ($k>1$ and $c_{1}>0$) completely
suppresses the MI if the CW amplitude is large enough,
$A_{0}^{2(k-1)}>|b|/ \left( kc_{1}\right) $.

\subsection{The model with dissipative nonlinearity}

In the case of $b=c_{1}=0$, Eq. (\ref{deq44}) takes a simple form,
\begin{equation}
w_{r}^{\prime \prime }+2\left[ kc_{2}/F(t)\right] w_{r}^{\prime }+\left[
q^{4}-4c_{2}^{2}k^{2}/\left( F(t)\right) ^{2}\right] w_{r}=0.  \label{deq81}
\end{equation}
Upon setting $\tau \equiv F(t)$, %$\tau=B_0^{-2 k}-2 k c_2 t$
Eq. (\ref{deq81}) reduces to
\begin{equation}
\tau ^{2}\frac{d^{2}w_{r}}{d\tau ^{2}}+\tau \frac{dw_{r}}{d\tau }+\left(
\frac{q^{4}}{4c_{2}^{2}k^{2}}\tau ^{2}-1\right) w_{r}=0,  \label{deq82}
\end{equation}
which is the Bessel equation. Its solutions are
\begin{equation}
w_{r}(\tau )=A_{1}J_{1}\left( \frac{q^{2}\tau }{2c_{2}k}\right)
+A_{2}Y_{1}\left( \frac{q^{2}\tau }{2c_{2}k}\right) ,  \label{deq83}
\end{equation}
where $A_{1,2}$ are arbitrary constants, and $J_1$, $Y_{1}$ are the 
Bessel functions
of the first and second kind, respectively. 
According to the known asymptotic properties of the
Bessel functions, this solution clearly shows transition to an oscillatory
behavior for large $\tau $. Notice that the case of $b=0$ and $c_{1}\neq 0$
is explicitly solvable too, through hypergeometric and Laguerre functions,
but the final expressions are rather cumbersome (and hence not shown here).

\section{Numerical results}

The above results indicate that the linearized equation governing
the evolution of the modulational perturbations can be explicitly
solved only in special cases. In order to analyze the general
case, we verified the long-time asymptotic predictions following
from Eq. (\ref{deq44}) by numerical simulations of this equation
in the case which is most relevant to the BECs. As mentioned above,
this one corresponds to $k=2$, while the other parameters in Eq.
(\ref{deq1}) are set, without loss of generality, to $b=-1$ and
$c_{1}=0$ (the latter condition is adopted since, typically
in the applications of interest, the conservative
part of the quintic term is less significant than the dominant
cubic term). Lastly, we fix $B_{0}=1$ in the CW solution
(\ref{F}), which is also tantamount to the general case, through a
scaling transformation of Eq. (\ref {deq1}), provided that $c_{2}$
is kept as a free parameter. Equation (\ref {deq44}) was solved using
a fourth-order Runge-Kutta integrator, with the time step
$dt=0.001$, and initial conditions $w_{r}(0)=0.05$ and $\dot{w}
_{r}(0)=0$.

\begin{figure}[tbp]
{\epsfig{file=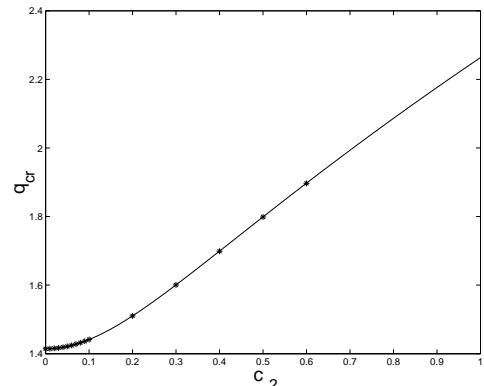, width=6.4cm,angle=0, clip=}}
\caption{The critical wavenumber $q_{{\rm cr}}$ [see Eq.
(\protect\ref{deq47})] of the band which gives rise to the
initial growth of modulational perturbations around the CW
solution. The solid line is the analytical prediction
(\protect\ref{deq47}), and the points show the numerical results. The
parameters are $B_{0}=1$, $b=-1$, $k=2$ and $c_{1}=0$.}\centering
 \label{fig1}
\end{figure}

Figure \ref{fig1} illustrates the critical wavenumber bounding the MI band
(of {\it transient} instability, if $c_2 \neq 0$)
as a function of the dissipation strength $c_{2}$. The solid line is the
analytical prediction given by Eq. (\ref{deq47}), which is compared with
points representing numerical data. The latter were obtained by identifying
the wavenumber for each value of $c_{2}$, beyond which there was no initial
growth of the perturbation $w_{r}$. Good agreement is between the
theoretical prediction and the numerical observations is obvious.

\begin{figure}[hbp]
%\epsfxsize=8.7cm 
%\epsffile{fd1n.ps}
%\epsfxsize=8.7cm 
%\epsffile{fd2.ps}
%\epsfxsize=8.7cm 
%\epsffile{fd3.ps}
%\epsfxsize=8.7cm 
%\epsffile{fd4.ps}
\centerline{
\epsfxsize=4cm 
\epsffile{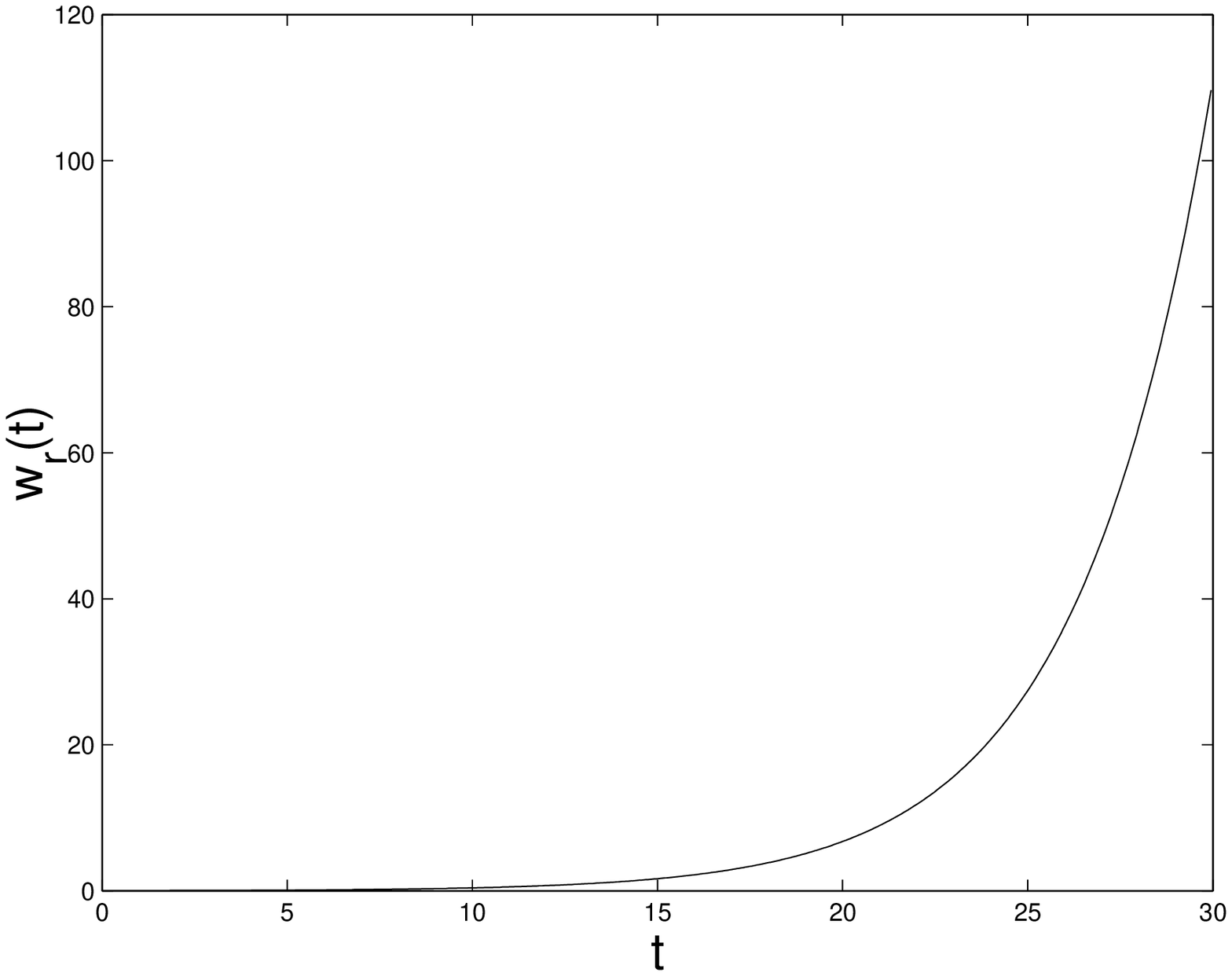}
\epsfxsize=4cm 
\epsffile{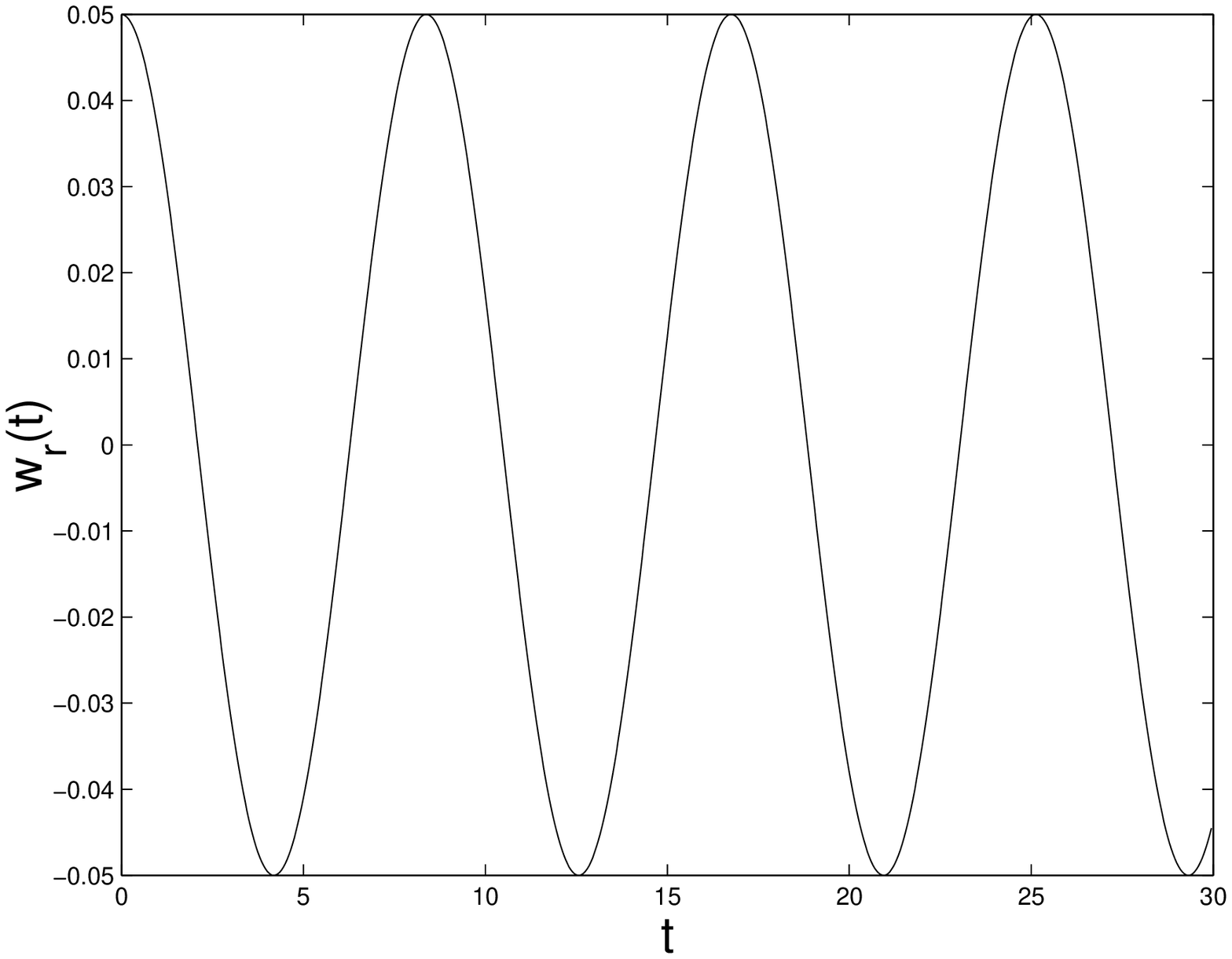}
}
\centerline{
\epsfxsize=4cm 
\epsffile{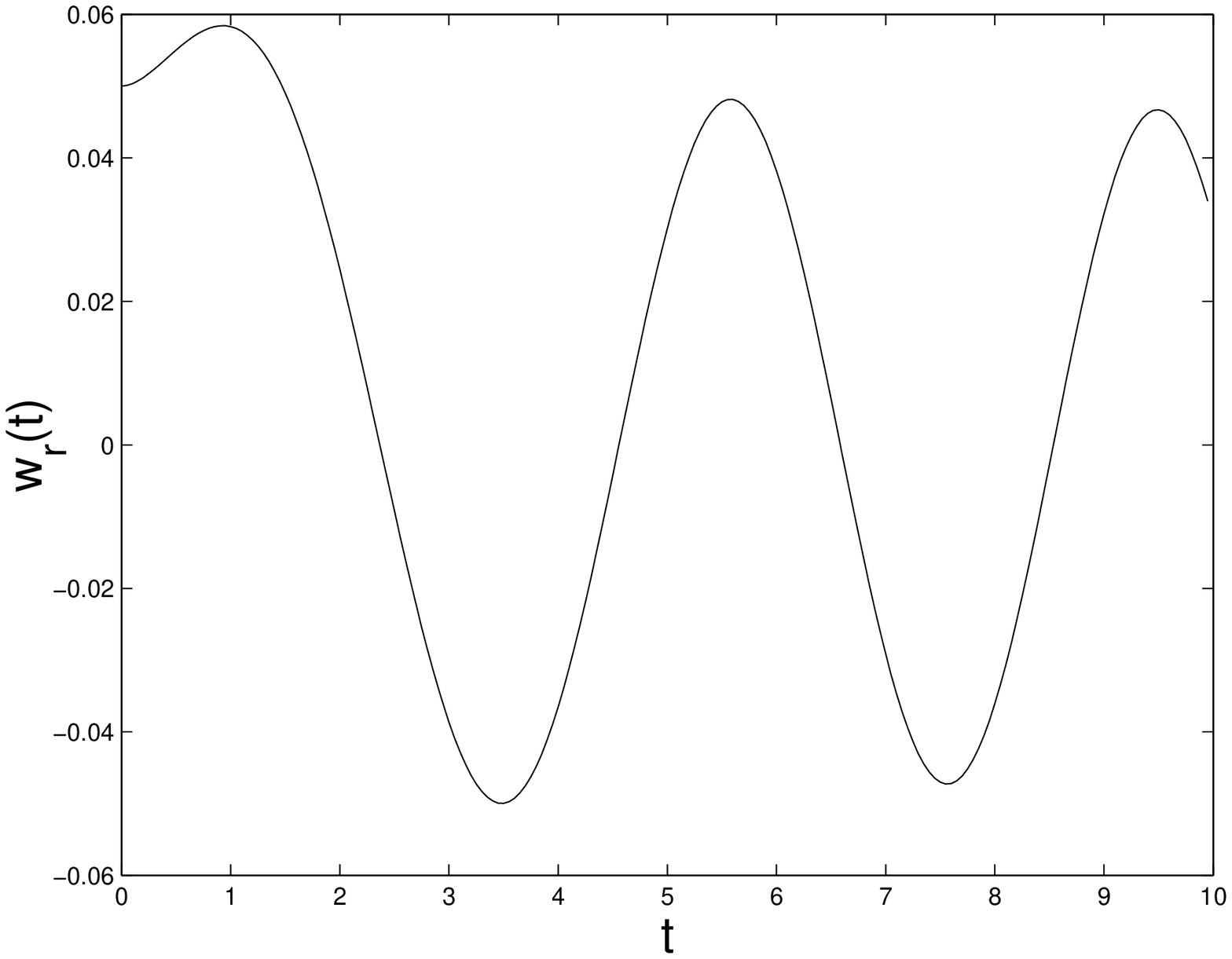}
\epsfxsize=4cm 
\epsffile{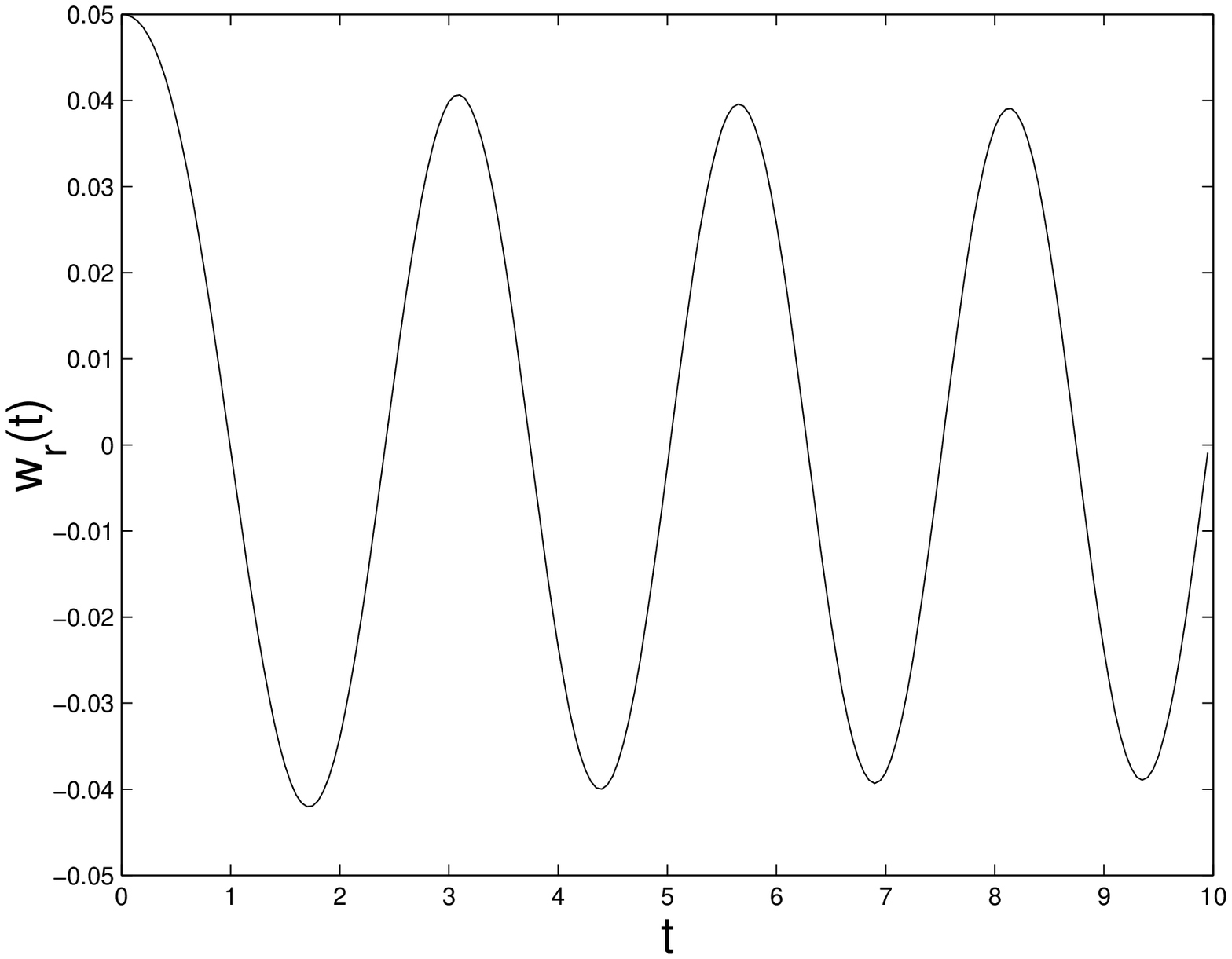}
}
\centerline{
\epsfxsize=4cm 
\epsffile{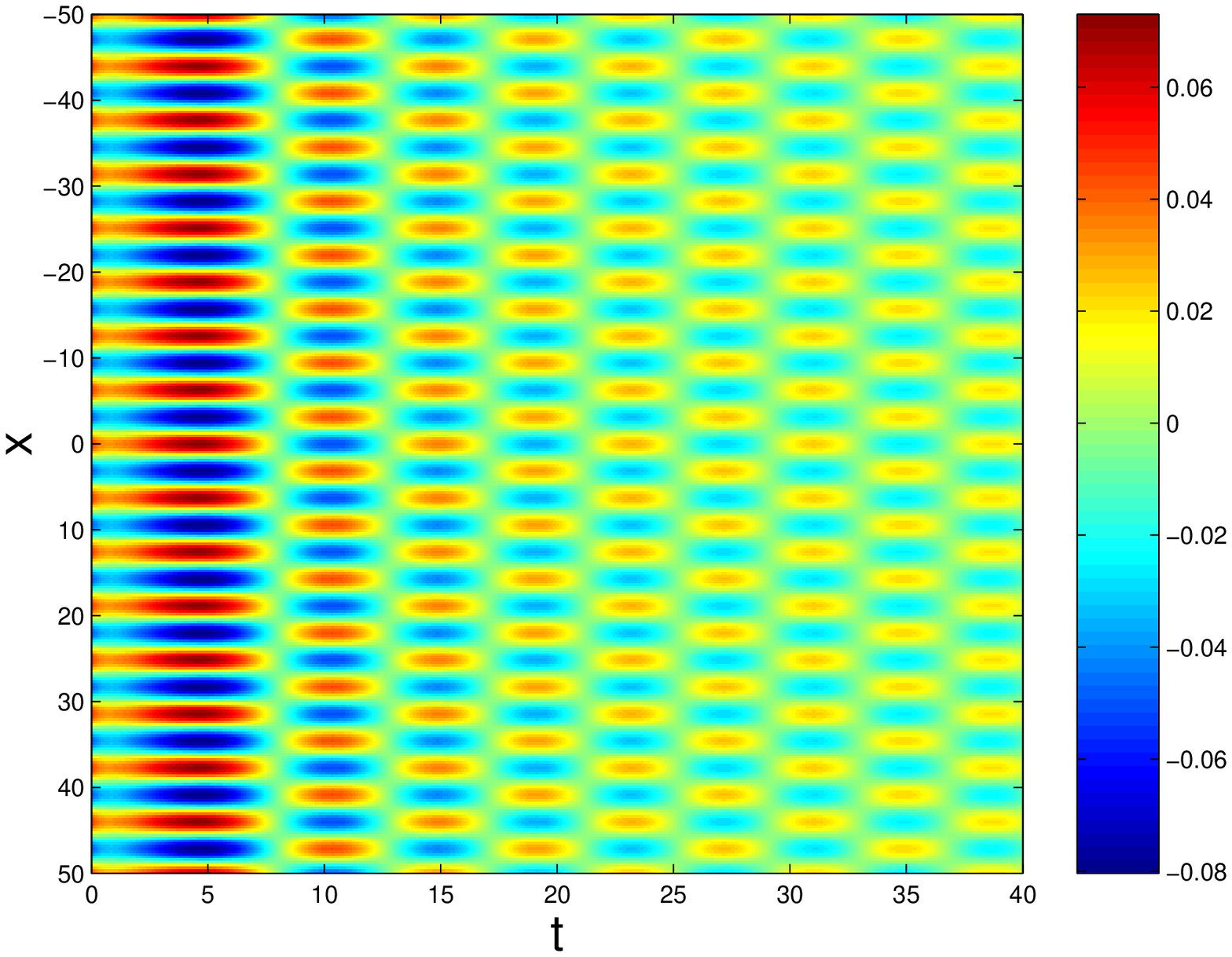}
\epsfxsize=4cm 
\epsffile{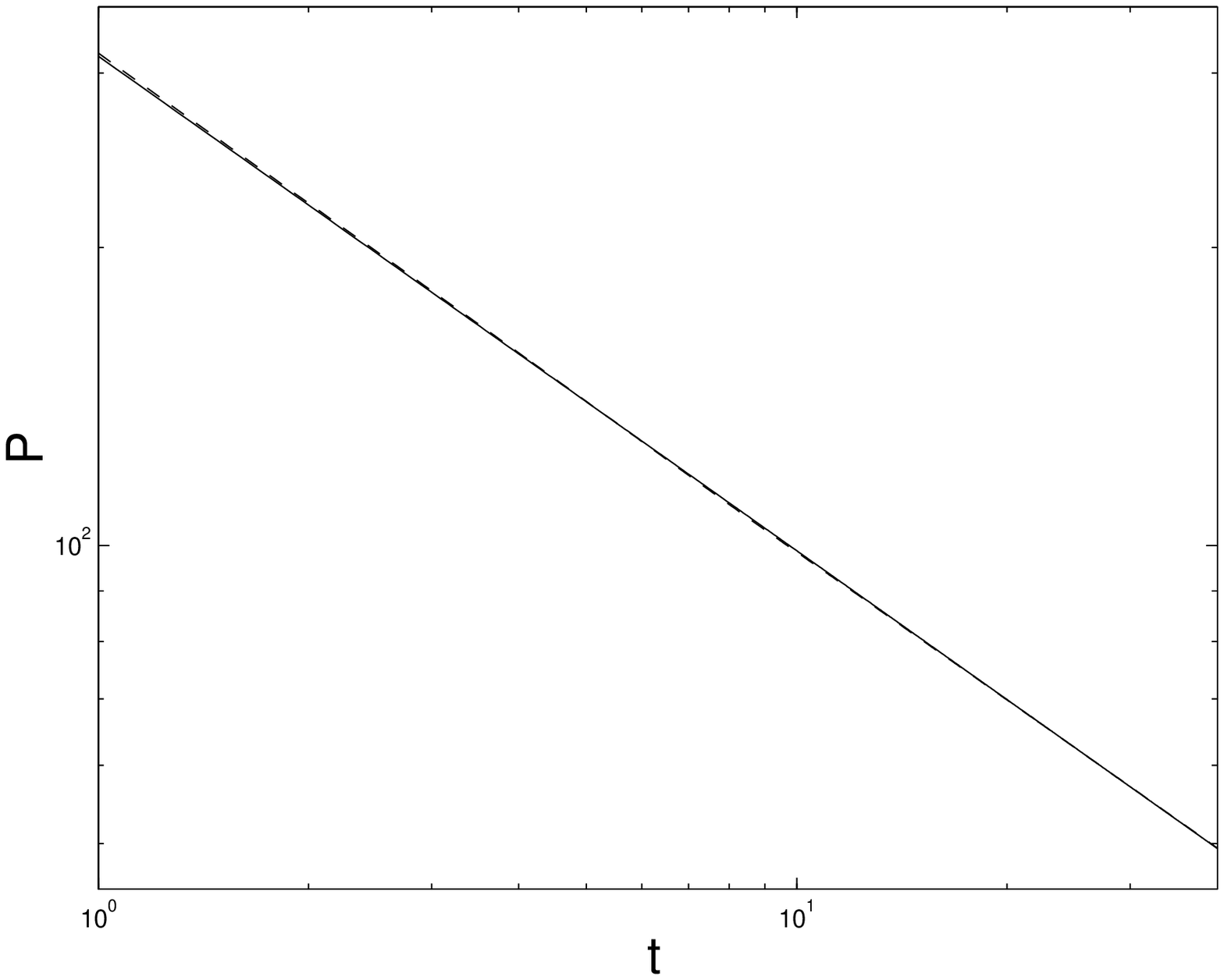}
}
\caption{
Top panels display
the evolution of the modulational perturbation, $w_{r}$, in the
absence of dissipation ($c_{2}=0$), for wavenumbers that do
($q=1.4$, left panel) or do not ($q=1.5$, right panel) belong to
the instability band. The middle panels display the same, but in
the presence of the nonlinear quintic dissipation ($|c_{2}|=0.3$),
for the perturbation wavenumbers $q=1.4$ (left panel) and $q=1.7$
(right panel).
The bottom panels show a typical result from the full simulation
of the Gross-Pitaevskii equation for $q=1$. The left panel shows
the space-time evolution of $(|u|-|u_0|)/|u_0|$, which 
evolves in a periodic form, whose amplitude depends on $w$.
We can see in the colorbar, the initial transient growth, followed
by the dissipative evolution. The right panel shows the best fit
(solid line)
and numerical (dashed line practically coinciding with the solid one)
time evolution of the norm $P=\int |u|^2 dx$. The best fit exponent
of $\log(P)$ vs. $\log(1+1.2 t)$ was found to be $-0.4994$ compared
with the theoretical exponent of $-0.5$ (according to Eqs. 
(\ref{deq3})-(\ref{F})).}
\label{fig2}
\end{figure}

Figure \ref{fig2} shows the difference in the evolution of
modulational perturbations whose wavenumber is chosen inside (left
panels) or outside (right panels) the instability band, in the
absence (top panels) and in the presence (middle and bottom panels) of the
quintic dissipation. The top left panel shows the case of $q=1.4$
for $c_{2}=0$, the top right illustrates the case of $q=1.5$ for
the same $c_{2}$, while the middle panels demonstrate the
evolution of $w_{r}$ for $|c_{2}|=0.3$, for $q=1.4$ (left panel)
and $q=1.7$ (right panel). The bottom panels show the evolution
of the norm in the GP equation (right panel) and the spatio-temporal
evolution of a function representing the perturbation, illustrating
(for $q=1$) the initial growth and eventual stabilization.

\section{Conclusions}

In this work we studied the modulational instability (MI) in equations of
the NLS type  in the presence of nonlinear dissipative
perturbations, motivated by the recent developments in the fields of
nonlinear optics, water waves and Bose-Einstein condensates (BECs). For
general power-law nonlinearities with both conservative and dissipative
parts, we analyzed the spatially uniform (CW) solutions and their
stability. We found that the dissipation generically suppresses the MI,
switching the growth of the perturbations into an oscillatory behavior. We
identified some special cases that can be analyzed exactly.
%, such as e.g., the
%corresponding to a purely conservative generalized NLS equation. 
Analytical
predictions based on the asymptotic consideration for $t\rightarrow \infty $
were compared with numerical simulations. A noteworthy prediction
(confirmed by numerical results) is that the band of wavenumbers for which
the transient MI occurs {\em expands} with the increase of the strength of
the dissipation.

As the MI is currently amenable to experimental observation in
BECs \cite {modin1}, it would be particularly interesting to
conduct experiments with different densities of the condensate.
With a higher density, three-body recombination effects and the
corresponding nonlinear dissipation are expected to be stronger,
affecting the onset of the MI as predicted in this work.

PGK gratefully acknowledges support from NSF-DMS-0204585, NSF-CAREER and
from the Eppley Foundation for Research. Harvey Segur is gratefully 
acknowledged for valuable discussions and for providing us a preprint
of Ref. \cite{segur}.

\end{multicols}

\end{document}